\documentstyle[12pt,aaspp4]{article}
\received{1996 September 9}

\begin{document}

\title{The Quadruple Gravitational Lens PG1115+080:\\ Time Delays and Models}
\author{Paul L. Schechter\altaffilmark{1},
Charles D. Bailyn\altaffilmark{2},
Robert Barr\altaffilmark{3},
Richard Barvainis\altaffilmark{4},
Christopher M. Becker\altaffilmark{1},
Gary M. Bernstein\altaffilmark{5},
John P. Blakeslee\altaffilmark{1},
Schelte J. Bus\altaffilmark{6},
Alan Dressler\altaffilmark{7},
Emilio E. Falco\altaffilmark{8},
Robert A. Fesen\altaffilmark{9},
Phillipe Fischer\altaffilmark{5,10},
Karl Gebhardt\altaffilmark{5},
Dianne Harmer\altaffilmark{11},
Jacqueline N. Hewitt\altaffilmark{1},
Jens Hjorth\altaffilmark{12},
Todd Hurt\altaffilmark{13},
Andreas O. Jaunsen\altaffilmark{14},
Mario Mateo\altaffilmark{5},
Doerte Mehlert\altaffilmark{15},
Douglas O. Richstone\altaffilmark{5},
Linda S. Sparke\altaffilmark{16},
John R. Thorstensen\altaffilmark{9},
John L. Tonry\altaffilmark{1},
Gary Wegner\altaffilmark{9},
Daryl W. Willmarth \altaffilmark{11} and
Guy Worthey\altaffilmark{5,10}}
\authoraddr{MIT Room 6-206, 77 Massachusetts Ave., Cambridge MA 02139}
\altaffiltext{1}{Department of Physics, Massachusetts Institute of Technology, Cambridge MA 02138}
\altaffiltext{2}{Department of Astronomy, Yale University, New Haven CT 06520-8101}
\altaffiltext{3}{Michigan-Dartmouth-MIT Observatory HC 04 P.O. Box 7520, Tucson AZ 85634}
\altaffiltext{4}{MIT Haystack Observatory, Westford MA 01886}
\altaffiltext{5}{Department of Astronomy, University of Michigan, Ann Arbor MI 48109}
\altaffiltext{6}{Department of Earth Atmospheric and Planetary Sciences, Massachusetts Institute of Technology, Cambridge MA 02139}
\altaffiltext{7}{Carnegie Observatories, Carnegie Institution of Washington, 813 Santa Barbara
Street, Pasadena CA 91101}
\altaffiltext{8}{Harvard-Smithsonian Center for Astrophysics, 60 Garden Street, Cambridge MA 02138}
\altaffiltext{9}{Department of Physics and Astronomy, Dartmouth College, Hanover NH 03755-3528}
\altaffiltext{10}{Hubble Fellow}
\altaffiltext{11}{National Optical Astronomy Observatories, P.O. Box 26732, Tucson AZ 85726}
\altaffiltext{12}{Institute of Astronomy, Madingly Road, Cambridge CB3 0HA, England}
\altaffiltext{13}{Department of Physics, University of California, Santa Barbara CA 93106}
\altaffiltext{14}{Nordic Optical Telescope, E-38700 Canarias, Spain}
\altaffiltext{15}{Universitaetssternwarte Muenchen, D-81679 Muenchen, Germany}
\altaffiltext{16}{Washburn Observatory, University of Wisconsin, Madison WI 53706}

\begin{abstract} 

Optical photometry is presented for the quadruple gravitational lens
PG1115+080.  A preliminary reduction of data taken from November 1995
to June 1996 gives component ``C'' leading component ``B'' by $ 23.7
\pm 3.4$ days and components ``A1'' and ``A2'' by $9.4$ days.
A range of models has been fit to the image positions, none of which
gives an adequate fit.  The best fitting and most physically plausible
of these, taking the lensing galaxy and the associated group of
galaxies to be singular isothermal spheres, gives a Hubble constant of
42 km/s/Mpc for $\Omega = 1$, with an observational uncertainty of
14\%, as computed from the $B-C$ time delay measurement.  
Taking the lensing galaxy to have an approximately E5 isothermal mass
distribution yields $H_0=64$ km/sec/Mpc while taking the galaxy to be a
point mass gives  $H_0=84$ km/sec/Mpc.  The former gives a particularly bad
fit to the position of the lensing galaxy, while the latter
is inconsistent with measurements of nearby galaxy rotation curves.
Constraints on these and other possible models are
expected to improve with planned HST observations.

\end{abstract}


\section{Introduction}

Among the most promising applications of gravitational lenses is
their use in determining cosmological distances (Refsdal 1964;
Blandford and Narayan 1992).  One measures fluxes for the components
of a multiply imaged variable object as a function of time and determines the
differential time delays.  Though straightforward in
principle, both the measurement of time delays and their
interpretation has proven difficult in practice.  Only recently have
measurements for 0957+561 (Haarsma {\it et al.} 1996a,b and
Kundi\'c {\it et al.} 1996 confirming the
result of Schild and Cholfin 1986) and their
interpretation (e.g.  Grogin and Narayan 1996) begun to
converge.  

The quadruply imaged $z=1.722$ quasar PG1115+080 (Weymann {\it et al.}
1980; see Figure 1) has many properties which recommend it as a
candidate for time delay measurements (Schechter 1996).  But while
PG1115+080 is known to vary (Vanderriest {\it et al.} 1986), and
though the flux ratios of the components appear to have varied
(Schechter 1996), as yet no time delay has been reported.  In the
following sections we report measurement of multiple time delays from a
preliminary reduction of data taken from November 1995 to June 1996.

\section{Observations and Reduction}

Direct $V$ filter CCD exposures were taken with the Hiltner 2.4-m
telescope, the WIYN 3.5-m telescope, the NOT 2.5-m telescope and the
Dupont 2.5-m telescope, always in multiples of at least three
exposures, typically 6 minutes each.  Frames were bias subtracted and
flatfielded and charged particle events were identified and flagged.
Six bright stars were used to obtain the
image scale and rotation.  

Photometry was carried out using methods nearly identical to those
used by Schechter and Moore (1993) for the lens MG0414+0534.  In the
present case positions for the lensing galaxy and the four quasar
images relative to each other were established from $V$ exposures
taken in 1994 with 0\farcs75 seeing, as were the shape parameters and the
galaxy flux.  The galaxy shape was convolved with the seeing
appropriate to each image.  Fixing the relative positions and the
galaxy shape and flux leaves six parameters to be determined from each
frame -- the four quasar fluxes and the overall position of the
system.

Quasar fluxes were obtained relative to the nearby star designated
``C'' by Vanderriest {\it et al.} (1986), which was used as an
empirical point spread function (PSF) template.  Fluxes for other
nearby stars relative to star ``C'' were also measured.  No attempt
was made to correct for color dependent airmass terms.

Fluxes were summed for the two brightest components, $A1$ and $A2$,
which are separated by only 0\farcs48.  Little is lost because the time
delay between $A1$ and $A2$ is expected to be hours.  Fluxes for each
multiplet of exposures were converted to relative magnitudes and
averaged together, and an rms scatter computed.  The median
scatters for components $A1+A2$, $C$ and $B$ were, respectively, 4, 7
and 13 millimagnitudes.  These were roughly 50\% bigger than the
formal errors in the fits.

The average magnitudes for the quasar components and for star ``B''
are plotted\footnote{These data can be found in tabular form at
http://arcturus.mit.edu/\char'176 schech/pg1115.html} in Figure 2.  They
include 30 nights of data from MDM Observatory and 2 from the NOT.
Multiple entries on the same night indicate exposures taken with
slightly different ``V'' filters.

Fifteen nights of data taken with the WIYN 3.5-m telescope were also
reduced but are not included.  The seeing at the WIYN and the
intranight consistency of these data were superb.  However the
night-to-night consistency was poor, and it has not proven possible
to bring the MDM data and the WIYN data into accord.  We expect that by
allowing for variation in the PSF across the chip we will ultimately be able to
combine the two data sets.

\section{Photometric Results}

In the light curves for components $A1+A2$, $C$ and $B$ shown in
Figure 2, the errors in the means, as computed from the intranight
scatter, are not much larger than the symbols.  The night-to-night
agreement is considerably worse; likewise the agreement of the shifted
lightcurves is much worse than expected from the intranight scatter.
On at least some nights the fluctuations, and those of star ``B'',
appear to be correlated (e.g. JD 24450088).  It seems likely that
these are due to errors in the adopted PSF.  Therefore in plotting the
quasar components in Figures 2 and 3, and in estimating time delays,
we have subtracted one half the magnitude of star ``B'' (taken with
respect to star ``C'') from our magnitudes for the quasar components
(also taken with respect to star ``C'').  This gives equal weight to
stars ``B'' and ``C'' as photometric references.

The amplitude for components $A1+A2$ appears somewhat smaller than
that for $C$ or $B$, but this is not unexpected.  Witt, Mao and
Schechter (1995) find that unresolved components suffer rms
fluctuations of 0.6 mag due to microlensing in quadruply imaged
systems.  If the time variable core is unresolved by the microlenses,
while most of the flux arises from a larger, resolved region, one
expects different fractional flux changes in the different components.

\section{Time Delay}

We have computed time delays and flux ratios for the $(A1+A2)-C$ and
$B-C$ pairs, using the method of Press, Rybicki, and Hewitt (1992),
but generalized so that the four parameters are fit to the three
light curves simultaneously.  Since the errors derived from the
intranight scatters are clearly too small, we have added to these,
in quadrature, additional errors of 4, 7, and 10 millimag for
components $A1+A2, C,$ and $B$ respectively.  The present analysis 
incorporates the implicit assumptions a) that the errors are uncorrelated
and b) that the fractional flux variations are the same for each component.

Under these assumptions, we find component $C$ leading component
$A1+A2$ by 9.4 days, and component $C$ leading component $B$ by $23.7$
days.  The best-fit $C/(A1+A2)$ flux ratio is 0.153 and the best-fit
$B/C$ flux ratio is $0.634$.  It follows that component $A1+A2$ leads
$B$ by 14.3 days and that the $B/(A1+A2)$ flux ratio is 0.097.  Figure
3 shows lightcurves for components $B$ and $C$, with the latter delayed.

A plot of the $\chi^2$ statistic for the $(A1+A2)-C$ and $B-C$ delays
shows a clean elliptical minimum, indicating an error for the $B-C$
delay which is 10\% larger than for $(A1+A2)-C$, with a positive
correlation coefficient of approximately 0.5.  This implies that
$B-(A1+A2)$ delay is correlated with the $B-C$ delay and
anticorrelated with the $(A1+A2)-C$ delay at roughly the same level.
The minimum value of $\chi^2$ was 146, with $\nu = 86$ degrees of
freedom.  Scaling errors to give $\chi^2/\nu = 1$ gives errors of
roughly $\sim 0.8$ days.

Monte Carlo simulations of the $B$ and $C$ components spanning a broad
range of delays and fitting only for the $B-C$ delay give an rms
scatter of $\pm 3.4$~days for the time delay, with only a weak
dependence on delay length.  These are considerably larger than those
obtained from the $\chi^2$ surface, for reasons discussed by Press,
Rybicki and Hewitt (1992).  Had we simulated all three components and
fit for the delays simultaneously the scatter might have been smaller.
For the present we adopt $\pm 3.4$~days as a best estimate for all of
the delays.

\section{Models, Distance and Hubble's Constant}

We have fit four models for the gravitational potential to the HST
positions (but not the fluxes) of Kristian {\it et al.} (1993) for the
purpose of transforming our time delay into a distance.  All but one
take the lensing galaxy to have a logarithmic 3-D potential, which
would produce flat rotation curve for circular orbits.  Model PMXS is
a point mass with external shear.  Model ISXS is an isothermal sphere
with external shear.  Model ISEP is an isothermal elliptical potential
(Blandford and Kochanek 1987).  The ISIS model, adopted by Hogg and
Blandford (1995) in modelling B1422+231, uses a second isothermal
sphere to provide shear. We identify this second isothermal with the
z=0.304 group of galaxies (Young {\it et al.} 1981; Henry and Heasley
1986) at approximately the same redshift as measured for the lensing
galaxy by Angonin-Willaime {\it et al.} (1993).

The dimensionless 2-D ``effective
lensing potentials'' $\psi$ are  given by 
\begin{mathletters}
\begin{eqnarray}
\psi_{PMXS}(\vec \theta)&=&b^2\ln r + {\gamma \over 2} r^2 \cos 2 (\theta - \theta_\gamma)\quad ,\\
 \psi_{ISXS}(\vec \theta) &=& br + {\gamma \over 2} r^2 \cos 2 (\theta - \theta_\gamma) \quad ,\\ 
 \psi_{ISEP}(\vec \theta) &=& br + \gamma br \cos 2 (\theta - \theta_\gamma) \quad {\rm and} \\  
 \psi_{ISIS}(\vec \theta) &=& br + b'r' \quad .  
\end{eqnarray}
\end{mathletters}
where $\vec \theta$ is angular position on the sky with respect to the
lensing galaxy with polar coordinates $r$ (an angle) and $\theta$, $b$
is the lens strength (also an angle), $\gamma$ is the dimensionless
shear, $\theta_\gamma$ gives the orientation of the shear (measured
from W to N, consistent with Kochanek 1991), $b'$ is the strength of a
second isothermal sphere and $r'$ is angular distance from this second
deflector.  The shear at any $r'$ for model ISIS is $\gamma = b'/2r'$.
Including the two unknown source positions ($\beta_W$ and $\beta_N$)
the first three models have 5 free parameters, while the last model
has 6 counting the P.A. and distance $d$ of the group from the lens.
Centering the coordinate system on the observed position of the galaxy
gives 8 observable quantities, the coordinates of the 4 images.

Parameter values and the source position were obtained by minimizing
residuals in the source plane.  The observed image positions were
projected back into the source plane, as were the position uncertainties
(which we took to be equal and circular).  The latter produce source
plane error ellipses with areas inversely proportional to the
magnification which are then used as a metric in minimizing the difference
between the model source position and the backward projections of the
image positions.  We take the uncertainty in the galaxy position to be
negligible in all but one model, labelled ISIS+, for which the
galaxy coordinates were taken as two additional free parameters.

Results for these fits are given in Table 1. The group strengths and
distances for the last two models are $b' = 2\farcs900$ and $2\farcs734$ and
$d = 14\farcs7$ and $13\farcs3$ respectively.  The strengths correspond to
velocity dispersions for the group of 383 and 372 km/s for $\Omega =
1$.  In the ISIS+ model the best position for the lensing galaxy was
only 11 mas from the position reported by Kristian {\it et al.}
(1993).

The derived model parameters given in Table 1 are similar to those
found by others (e.g. Kochanek 1991; Narasimha and Chitre 1992;
Keeton, Kochanek and Seljak 1996) for the same system.  The elliptical
potential (ISEP) fit least well, and would require an E5 or flatter
mass distribution.  The double isothermal (ISIS) fit significantly
better than the isothermal with external shear (ISXS), even when the
position of the ``tidal'' isothermal was fixed at the flux weighted
centroid of the neighboring galaxies (c.f. Young {\it et al.} 1981) at
P.A. -116.4$^\circ$ E of N and $ d = 19\farcs1$.  This corresponds to a
shear at $\theta_\gamma = 63.6^\circ$ N of W.  The goodness of fit,
the close positional coincidence, and the plausible value of the group
velocity dispersion are strong arguments in favor of the ISIS model.
Following Narayan and Bartelmann (1996), time delays $\tau$ are given
by
\begin{equation}
\tau(\vec
\theta) = {(1+z_l) \over c} {D_l D_s \over D_{ls}}
\left[{1 \over 2}(\vec \theta - \vec \beta)^2 - \psi(\vec \theta)\right] \quad ,
\end{equation}
where $\vec\beta$ is the source position, the distances $D$ are angular
diameter distances, and where the subscripts $l$ and $s$ refer to lens
and source, respectively.  The left and right hand terms inside the
square brackets are, respectively, the geometric and gravitational
components of the time delay.  Also shown in Table 1 are
time delays in days computed for $\Omega = 1$ and $H_0 = 100$ km/s/Mpc, taking the
galaxy redshift to be the same as the group redshift found by Henry
and Heasley (1986), $z = 0.304$, and the source redshift to be $1.722$.
For these values we find
\begin{displaymath}
{(1+z_l) \over c} {D_l D_s \over D_{ls}} = 30.49 \; {\rm days \; arcsec^{-2}}.
\end{displaymath}
Models with $(\Omega, \lambda) = (0.1, 0.0)$ predict time delays longer by
a factor of 1.074; those with  $(\Omega, \lambda) = (0.1, 0.9)$ predict
delays longer by 1.038.

Model PMXS predicts longer time delays because the lensing galaxy is
more centrally concentrated.  Wambsganss and Paczynski (1994) have
noted this effect and emphasize the difficulty in constraining the
degree of central concentration with only the positions of 4 images.
Model ISEP predicts longer time delays because the source must lie
near the ``diamond caustic'' to produce the two brighter images.  The
diamond caustic for an ISEP model is roughly twice that of an ISXS
model with the same shear, forcing the source further off center and
increasing the difference in arrival time.  A change in position of
the lens of 11 mas produces a 3\% change in the $B-C$ time delay.
Kristian {\it et al.} (1993) quote a 50 mas uncertainty for that
position.

Comparison of the observed $B - C$ delay of 23.7 days with the
predictions of Table 1 gives $H_0 =$ 84, 44, 64, 41 and 42 km/s/Mpc
for the PMXS, ISXS, ISEP, ISIS and ISIS+ models, respectively.  The
14\% uncertainty in the time delay gives a 14\% uncertainty in each of
these numbers.  Interpolating between the PMXS and ISXS models, we
estimate that taking the leading term in equations (1b-d) to vary as
$r^{1\pm0.2}$ produces an additional 14\% uncertainty in the Hubble
constant.

In all of the models the $A-C$ delay is consistently a factor of 1.5
larger than the $B-A$ delay, while the measured $A-C$ delay is a factor
of 1.5 {\it smaller}.  While this discrepancy is only at the $2
\sigma$ level for our adopted uncertainties, it is nonetheless
cause for concern.  We believe that the $B-C$ delay is less subject to
systematic error than either of the shorter delays because of the
larger number of samples per time delay.

\section{Prospects for Improvement}

There are many ways in which our time delay and distance estimate
might be improved upon.  As described above, the intranight scatter in
the photometry is considerably closer to the photon limit than the
night-to-night scatter.  We expect that a new reduction with improved
algorithms, a better profile for the lensing galaxy and perhaps with
better flatfielding will improve the scatter and decrease the
uncertainty in the time delay.  Better algorithms may allow the
inclusion of roughly 20 nights of WIYN data which fill many of the
gaps and have not been included in the present analysis.  Finally, not
all of our data have been reduced.  Some 25 nights of data taken with
the Hiltner telescope, the NOT, and DuPont telescope await reduction.
Further monitoring would likewise reduce the uncertainties in the time
delays and help to clarify the discrepancy between the observed
intermediate delays and those predicted by our models.  New HST data
to be obtained during Cycle 6 is expected to reduce the uncertainty in
the lens position and give an ellipticity for the lensing galaxy.

\acknowledgments
We thank M. Pierce for obtaining observations of PG1115+080 on our
behalf.  PLS thanks G. Canalizo and E. Fulton for their early efforts
in PSF fitting and lens modelling, and P. Steinhardt for pointing out a
small numerical error in an early version of the manuscript.  DOR
thanks the Institute for Advanced Study for its hospitality.

\clearpage

\figcaption[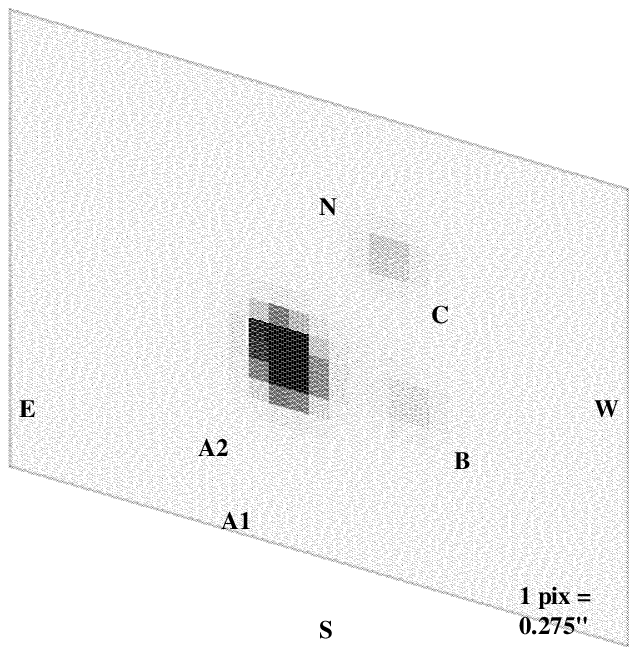]{An I filter direct image of PG1115+080 showing
QSO components $A1, A2, B,$ and $C$.  Stars ``B'' and ``C'' of Vanderriest
{\it et al.} (1986) are outside the field. \label{fig1}}

\figcaption[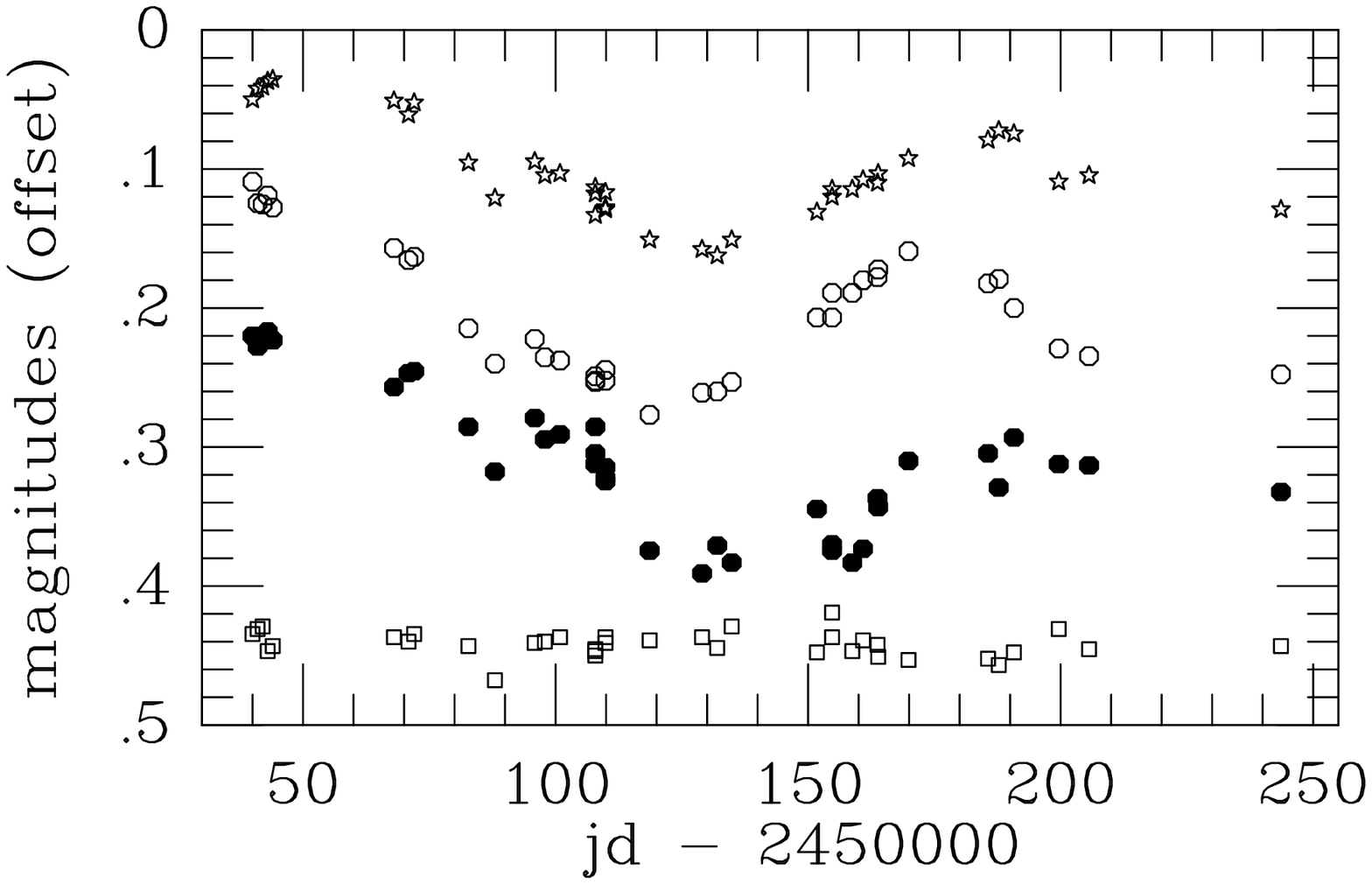]{Unshifted light curves for the PG1115+080 components
components $A1+A2$ ($\star$), $C$ ($\circ$), and $B$ ($\bullet$) and for
reference star ``B'' ($\Box$).
\label{fig2}}

\figcaption[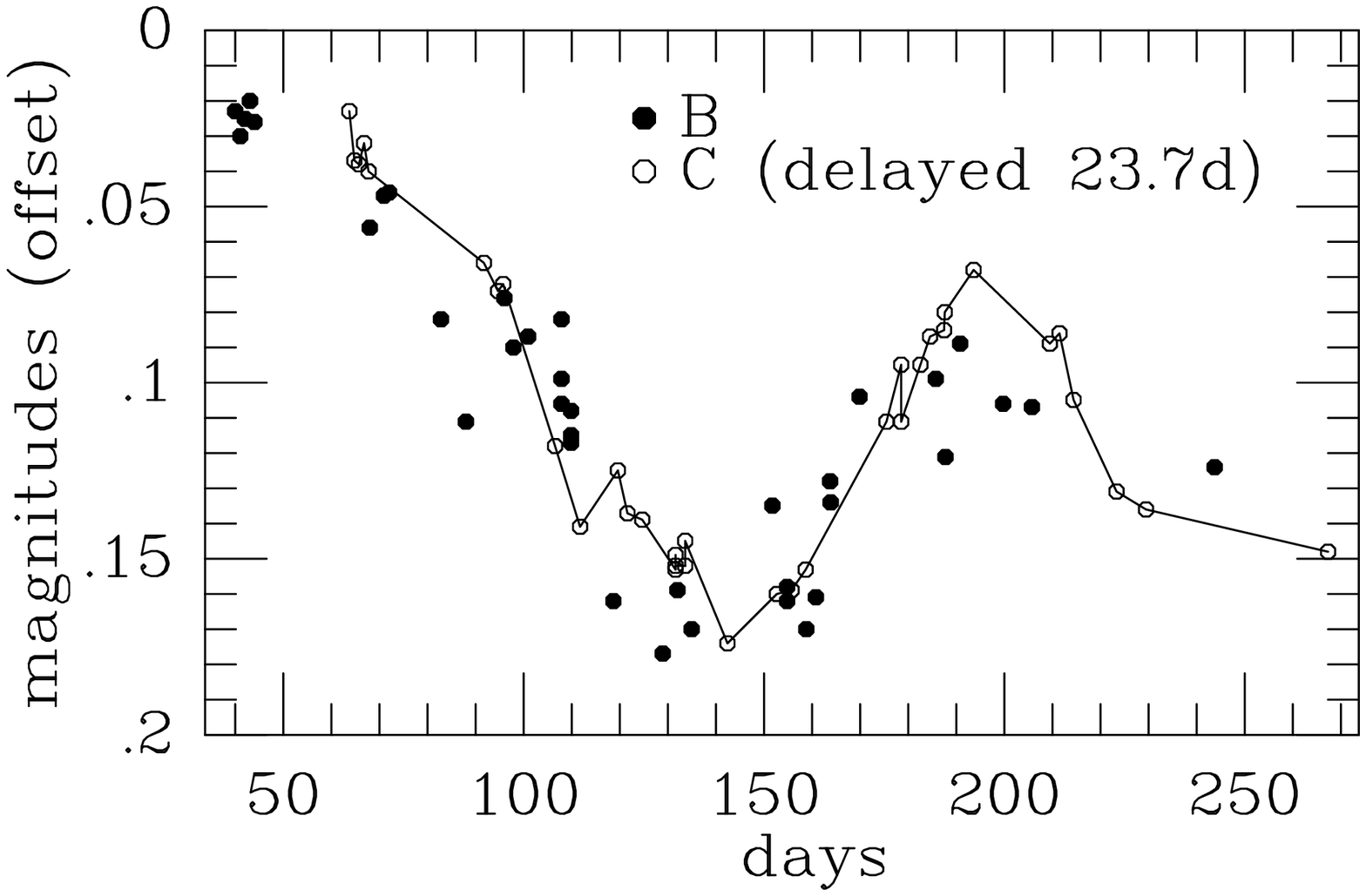]{Light curve for component $C$ ($\circ$) shifted by
23.7 days to match that of component $B$ ($\bullet$).
\label{fig3}}

\clearpage

\begin{table}
\caption{Model Parameters and Time Delays for Lens Models}
\begin{tabular}{lrrrrrrrrrr}
model & b & $\beta_W$ & $\beta_N$ & $\gamma$ & $\theta_\gamma$ &
 A1-C & A2-A1 & B-A2 & B-C & $\chi^2/DOF$ \\
        &  ('') & ('') & ('') & ( ) & ($^\circ$) & (d) & (d) & (d) & (d) & \\
\tableline
PMXS  & 1.137 & -0.047 &  0.205 & 0.198 & 65.8 & 12.5 & 0.13 & 7.3 & 19.9 & 135/3\\
ISXS  & 1.144 & -0.025 &  0.108 & 0.103 & 65.8 &  6.6 & 0.07 & 3.8 & 10.4 & 104/3\\
ISEP  & 1.164 & -0.022 &  0.151 & 0.079 & 65.7 &  9.7 & 0.43 & 5.0 & 15.1 & 204/3\\
ISIS  & 1.033 &  2.598 & -1.124 & 0.099 & 65.1 &  5.6 & 0.06 & 4.0 &  9.7 &  10/2\\
ISIS+ & 1.026 &  2.446 & -1.056 & 0.103 & 64.9 &  5.7 & 0.07 & 4.2 & 10.0 &   0/0\\
\end {tabular}
\end{table}

\clearpage

\begin{figure}[h]
\vspace{3.0 truein}
\includegraphics{pg1115_blowup.ps}
\end{figure}

\clearpage

\begin{figure}[h]
\vspace{7.0 truein}
\includegraphics{first_half.ps}
\end{figure}

\clearpage

\begin{figure}[h]
\vspace{7.0 truein}
\includegraphics{second_half.ps}
\end{figure}

\end{document}